\documentclass[preprint,groupedaddress,amsmath,amssymb,prb,lengthcheck,showpacs]{revtex4-1}

\usepackage[T1]{fontenc}
\usepackage[latin1]{inputenc}  
\usepackage{graphicx} 
\usepackage{hyperref}  
\usepackage{color} 
\usepackage{upgreek}
\usepackage{amsmath}
\usepackage{amssymb}
\definecolor{darkred}{rgb}{0.4,0,0} 
\definecolor{darkgreen}{rgb}{0,0.2,0} 
\definecolor{darkblue}{rgb}{0,0,0.4} 
\hypersetup{colorlinks=true, breaklinks=true, linkcolor=darkblue, menucolor=darkblue, urlcolor=darkblue}
\hypersetup{colorlinks,linkcolor=darkblue,filecolor=black,urlcolor=darkred,citecolor=darkblue} 
\hypersetup{colorlinks,linkcolor=darkblue,filecolor=darkgreen,urlcolor=darkred,citecolor=darkblue} 

\newcommand{\nm}{\ensuremath{\,\mathrm{nm}}}

\newcommand{\mum}{\ensuremath{\,\mathrm{\upmu{} \mathrm{m}}}}

\newcommand{\GHz}{\ensuremath{\,\mathrm{GHz}}}

\newcommand{\Apm}{\ensuremath{\,\mathrm{A}/\mathrm{m}}}
\newcommand{\Jpkm}{\ensuremath{\,\mathrm{J}/\mathrm{m}^3}}

\newcommand{\degree}{^\circ}

\newcommand{\be}{\begin{equation}}
\newcommand{\ee}{\end{equation}}
\newcommand{\chii}{\ensuremath{\,\mathrm{\,{\chi}}\,}}
\newcommand{\chiii}{\ensuremath{\,\mathrm{\,{\boldsymbol{\chi}}}\,}}

\newcommand{\dm}{\ensuremath{\vec{m}\,}}
\newcommand{\M}{\ensuremath{\vec{M}\,}}
\newcommand{\dmz}{\ensuremath{m_z\,}}
\newcommand{\hrf}{\ensuremath{\vec{h}_{RF}\,}}
\newcommand{\hrfamp}{\ensuremath{\,h_0\,}}
\newcommand{\hst}{\ensuremath{\vec{H}\,}}

\newcommand{\ts}{\ensuremath{t_{s}\,}}

\let\baraccent=\= 
\renewcommand{\=}[1]{\stackrel{#1}{=}} 

\begin{document}

\title{Continuous wave approach for simulating Ferromagnetic Resonance in nanosized elements}

\author{K. Wagner}
\email{k.wagner@hzdr.de; Corresponding author}
\author{S. Stienen}
\affiliation{Helmholtz-Zentrum Dresden-Rossendorf, \\ Institute of Ion Beam Physics and Materials Research, \\ Bautzner Landstraße 400, 01328 Dresden, Germany}

\author{M. Farle}
\affiliation{Faculty of Physics and Center for Nanointegration (CeNIDE), \\ University of Duisburg-Essen, Lotharstr. 1, 47057 Duisburg, Germany}

\date{\today}

\begin{abstract}

We present a numerical approach to simulate the Ferromagnetic Resonance (FMR) of micron and nanosized magnetic elements
by a micromagnetic finite difference method. In addition to a static magnetic field a linearly polarized oscillating magnetic field
is utilized to excite and analyze the spin wave excitations observed by Ferromagnetic Resonance in the space- and time-domain.
Our continuous wave approach (CW) provides an alternative to the common simulation method, which uses a pulsed excitation of
the magnetic system. 
It directly models conventional FMR-experiments and permits the determination of the real and imaginary part of the complex dynamic susceptibility without the need of post-processing.
Furthermore not only the resonance fields, but also linewidths, ellipticity, phase relations and relative intensities of the excited spin wave modes in a spectrum can be determined and compared to experimental data. 
The magnetic responses can be plotted as a function of spatial dimensions yielding a detailed visualization of the spin wave modes and
their localization as a function of external magnetic field and frequency. This is illustrated for the case of a magnetic micron sized
stripe.
\end{abstract}


\maketitle

\section{Introduction}\label{Introduction}
The detailed understanding of spin wave spectra of magnetic
micro- and nanostructures and their magnetization dynamics
has found increasing interest from both fundamental and applied
points of view for example in spin caloritronics and spin
torque phenomena \cite{Magnonics,BockMagnonics,spinwaveSpintorque,Spincalorics}. A powerful tool to investigate
these spin wave spectra experimentally is the Ferromagnetic
Resonance (FMR) detected in the frequency domain \cite{Farle2013,Lindner}. However in most cases the obtained FMR-spectra are complex
in nature featuring several -often overlapping- resonances
and require theoretical descriptions of the nanostructured magnetic
systems to extract quantitative information. Micromagnetic
simulations of the FMR can be used to model those systems
and provide additional information on the character of the
observed magnetic excitations as well as their dependence on
magnetic parameters, geometries, confinement effects or charge
currentss \cite{McMichael,Venkat}. This is especially of interest when the complex geometries and interactions of the nanoscale ferromagnet aggravate quantitative analytical approaches. 

Here we present a finite difference method utilizing a homogeneous oscillating magnetic field to simulate FMR spectra corresponding to experiments.
In addition we show how to further analyze the spectra by visualizing the spatial distribution of the magnetic excitations. 
We start by describing the problem definition, initialization and recorded data of the simulations in section \ref{method}. Subsequently in section \ref{Analysis} the derivation of FMR-spectra is described in detail as well as determination of the FMR fields, linewidth and ellipticity of the resonances. In section \ref{modeprofile} we investigate the resonances contained in the spectra in terms of spin waves and spatial variations.

\section{Method}\label{method}

The micromagnetic simulations presented here, are based on the public domain 3D-OOMMF (Object Oriented MicroMagnetic Framework)\cite{OOMMF} solver. This finite difference software solves numerically the Landau-Liftshitz equation \cite{LLG} (LLE)
\begin{equation}
\frac{d\M}{dt}=-\gamma\left(\M\times \vec{H}_{\textrm{eff}}\right)-\frac{\gamma\cdot\alpha}{M_{s}}\left(\M\times\left(\M\times \vec{H}_{\textrm{eff}}\right)\right)
\end{equation}
where $\gamma$ is the gyromagnetic ratio, $\alpha$ the Gilbert damping constant and $\vec{H}_{\textrm{eff}}$ the effective field.
As time evolver a Runge-Kutta method is used. Further details on the implemention can be found in Ref. \hyperlink{target}{5}. Our approach to simulate FMR-experiments can be split into three different steps: 1. relaxation 2. transient phase 3. dynamic equilibrium. 

For initialization the spatial dimensions of a nanostructured ferromagnetic system are defined by a grid of equal rectangular cells. All cells are assigned with an identical magnetization vector $\M$, located at the center of the cell. In order to simulate the external field of FMR-experiments, a static magnetic field $\hst$ is applied to the system. To obtain the static magnetic ground state $d\M/dt=0$ a relaxation simulation is performed, without applying any excitation. So the motion of $\M$ damps out and $\M$ will reorient to an equilibrium direction, given by the local effective field. For faster convergence the precession term in the LLE may be switched off and the damping constant $\alpha$ set large. As stopping criterion for the simulation typically values of $d\hat{u}/dt<0.001\ ^\circ/\text{ns}$ ($\hat{u}$ is the unit vector of $\M$) are chosen, to achieve the quasi static state, which is used as the inital state for the subsequent FMR-simulations.
 
In addition to the static field $\hst$ a linearly polarized oscillating magnetic field $\hrf(t)=\vec{\hrfamp}\cdot \sin(\omega\cdot t)$ is added for continous excitation of the magnetization (CW). $\hrf$ is uniform over all cells, oriented perpendicular to the static field and corresponds to the magnetic microwave field used in conventional FMR-experiments: $\hrfamp=398\ \text{A/m}\ \hat{=}\ 0.5\ \text{mT}$ (satifying the relation $\hrfamp<<\hst$).

Due to the interaction with $\hrf$ a driving torque is exerted on the magnetization. After a transient phase the magnetization reaches a dynamic equilibrium, precessing around the effective field with the angular frequency $\omega$. In this state $\hrf$ transfers power to the magnetic system to compensate dissipation, induced by damping. To study the dynamic equilibrium and discard transient effects, a fixed time period is simulated, without generating data for analysis. This time period is given by $\ts=2\pi\cdot N/\omega$, using an integer oscillation number $N$ of $\hrf$ (details described in section \ref{Analysis}). $N$ is typically set between 40 and 60 depending on the magnetic parameters (e.g. damping constant $\alpha$). This provides a constant precession amplitude between consecutive oscillation cycles of the magnetization with a deviation of less than $0.02\ \%$. When the simulation time reaches $\ts$ the actual parameters (like magnetization $\M$, oscillating field $\hrf$, static field $\hst$,...) are stored for each following iteration step of the time evolver. This process continues for one further cycle of the oscillating field ($N+1$) and consists of at least 1000 iteration steps, which provides a time resolution in the ps regime. By analyzing these data, it is possible to follow the precession trajectory of the cell specific magnetization in the space- and time-domain.

\section{Analysis of the simulation}\label{Analysis}
\begin{figure}[h!tbp]
\includegraphics[width=0.9\columnwidth]{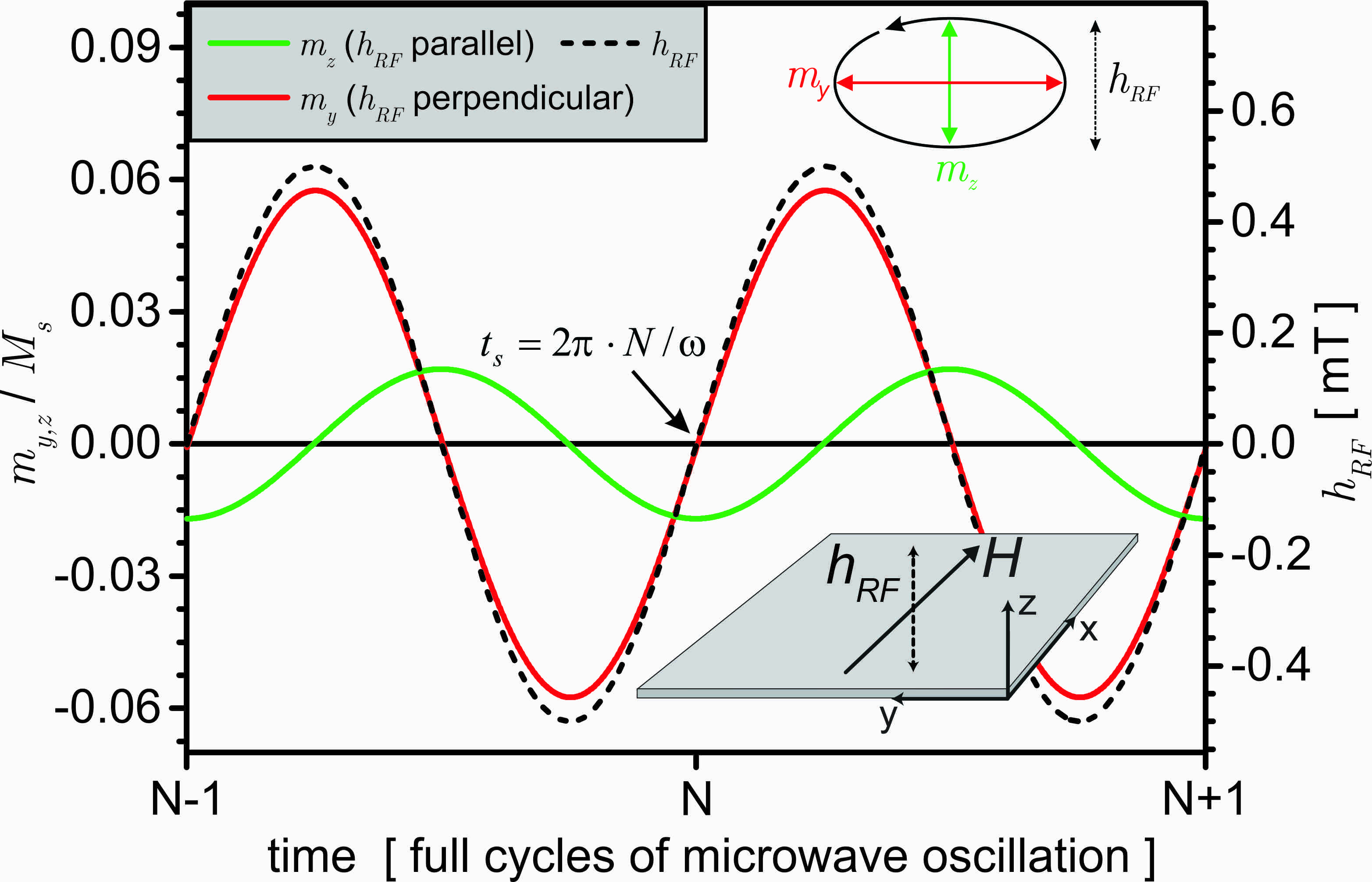}
\caption{Micromagnetic simulation of the time dependent response of the dynamic magnetization driven by an external dynamic field $\hrf$ for an infinite thin film in the xy-plane. The normalized dynamic magnetization in the yz-plane (solid lines) are shown together with $\hrf$ (dashed line) oriented along the film-normal in the z-direction. The static field $\hst$ is oriented in the film plane (x-direction) and is chosen to match the resonance condition.} \label{fig1}
\end{figure}

We now describe the analysis of the simulated data for the dynamic equilibrium. In the linear response regime discussed here the dynamic magnetization $\dm(t)$ of each cell exposed to the oscillating field is described by the dynamic susceptibility-tensor $\chiii=\chiii'+i\chiii''$: 

\begin{equation}
\dm(t) = \chiii \,\hrf(t) = \chiii \, \vec{\hrfamp} \exp{-i(\omega t - \pi/2)}
\label{eq-magneticresponse} 
\end{equation}

Where we have used a complex representation of the susceptibility and the applied sinusoidal varying magnetic field $\hrf$. $\chiii$ is a $3 \times 3$-tensor with elements $\chii_{ij}$.

To illustrate the motion of $\dm$ in respect to the $\hrf$, the simulated time dependences for the case of an infinite film spanning the xy-plane are shown in fig. \ref{fig1}. $\hrf$ is oriented in the z-direction (out of plane) and $\hst$ in the x-direction (in plane), respectively. The displayed time dependent components of the oscillating dynamic magnetization (solid lines) lie in the yz-plane driven by $\hrf$ (dashed line).
The oscillation of $\dm$ and its components $m_{y,z}$ is described by its amplitude $A_{y,z}$, frequency $\omega$ and phase relation $\phi_{y,z}$ in respect to $\hrf$. Note that the driven component $m_z$ exhibits a phase-shift of $90 \degree$ to the oscillating field as expected for a resonantly driven system. The precessional motion of the magnetization in equilibrium as well as the ellipticity of its trajectory can be readily observed by the $90 \degree$ phase shift between the dynamic components $m_{y,z}$ and the ratio of their differing maximal amplitudes.

The simulated FMR-spectra, which can be quantitatively compared to experimental ones are derived as described in the following. The power $P$ absorbed by the magnetic system from $\hrf$ and consequently the FMR-signal $S$ is proportional to the diagonal element $\chii_{zz}''$ of the imaginary part of the susceptibility \cite{Vonsovskij}:

\be
S \propto P \propto \chii_{zz}''
\ee

To determine $\chii_{zz}''$ from the simulation it is sufficient to analyze the component of $\dm$ parallel to $\hrf$ (in this case $m_z$) after a complete cycle of oscillations as given for the time $t_s$ (see also section 2). This can be seen when considering the observable real part of $\dm$ in equation 2:

\begin{eqnarray}\label{eq3}
\dmz && = \Re \left( \chii_{zz} \hrfamp \exp{(-i (\omega t - \pi/2))} \right)
\end{eqnarray}

Inserting $t_s$ ($\omega\ts = 2 \pi\,N$) yields:

\begin{eqnarray}\label{eq4}
\dmz (\ts) && = -\chii''_{zz}\,\hrfamp
\end{eqnarray}

Hence, a proportional FMR-Signal ($\chi_{zz}''$) can be simulated by directly monitoring $-\dmz (\ts) / \hrfamp$ without the need for extensive post-processing. (By a similar logic) the real part of the susceptibility $\chi_{zz}'$ - and therefore the complete $\chi_{zz}$ - can be retrieved from the simulation by extracting $m_z$ at the maximum of $\hrf$.

\begin{figure}[htbp]
\includegraphics[width=0.9\columnwidth]{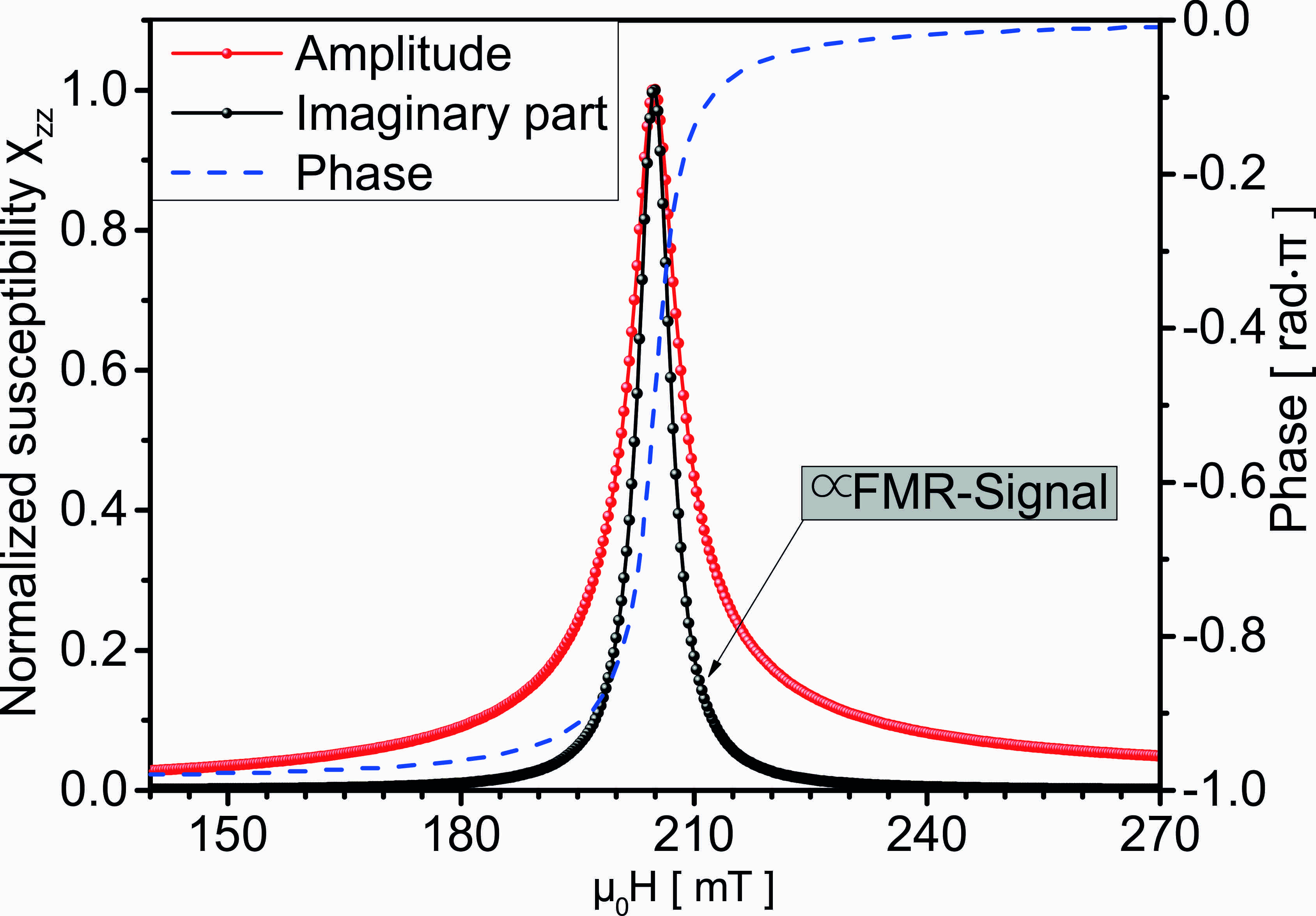}
\caption{Micromagnetic simulation of amplitude, phase and imaginary part of the susceptibility for an infinite thin film for different static fields and a driving frequency of $15 \GHz$. The magnetic resonance exhibits the classical hallmarks of a driven oscillator, where the imaginary part is proportional to the experimentally detected FMR-Signal.} \label{fig2}
\end{figure}

Fig. \ref{fig2} shows the simulated external field dependent amplitude, phase and imaginary part of the susceptibility for an infinite thin film with the magnetic parameters: exchange constant $A = 13\cdot 10^{-12}\Jpkm$, saturation magnetization $M_s = 8.3 \cdot 10^{6} \Apm$, g-factor $g=2.12$, Gilbert damping parameter $\alpha=5\cdot 10^{-3}$.

The magnetic response shows the typical hallmarks of a driven oscillator in respect to phase and amplitude and a lorentzian absorption curve \cite{Poole}. 
This enables one to determine the resonance positions as well as their linewidth and relative signal strength. 
In the case of multiple resonances a decomposition of the resultant spectra (experiment or simulation) into lorentzian absorption lines is needed to obtain those quantities.
Subsequently this can be compared to complex experimental results as for example for the case of magnetic micron sized stripes \cite{Banholzer,Schoeppner,Zheng}.

\section{Characterizing the magnetic excitations}\label{modeprofile}
\begin{figure}[htbp]
\includegraphics[width=0.9\columnwidth]{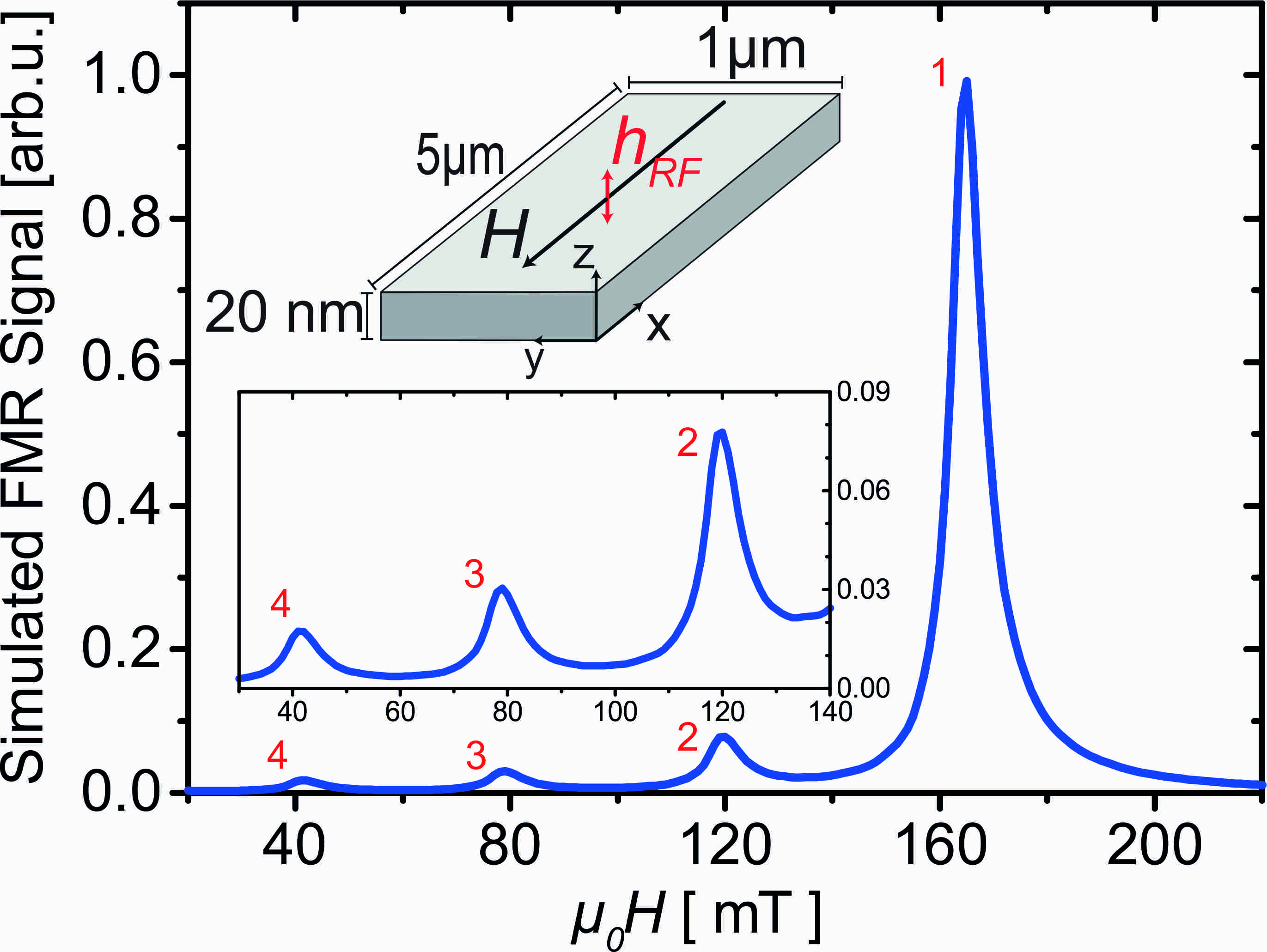}
\caption{Micromagnetic simulation of the susceptibility $\chii''_{zz}$ for a $5 \mum \times 1 \mum \times  20 \nm$ Permalloy stripe and a frequency of $14\GHz$. In the confined system multiple magnetic resonances are observed (labeled from  1 to 4)} \label{fig3}
\end{figure}

To further investigate the magnetic resonances in the simulated FMR spectra the amplitude, phase and imaginary part of the susceptibility may be analyzed for each cell of the magnetic system individually. The retrieved spatial distribution of the magnetic response often helps to identify spinwave like resonances, their wavevector or a localized character of the excitations. Here we exemplarily simulate a FMR Spectrum of a $5 \mum \times 1 \mum \times  20 \nm$ stripe at $14\GHz$. As magnetic parameters we choose: exchange constant $A = 13\cdot 10^{-12}\Jpkm$, saturation magnetization $M_s = 8.3 \cdot 10^{6} \Apm$, g-factor $g=2.13$, Gilbert damping parameter $\alpha=7\cdot 10^{-3}$. The simulated FMR-Signal together with the stripe geometry and field orientation is shown in fig. \ref{fig3}.

\begin{figure}[htbp]
\includegraphics[width=0.9\columnwidth]{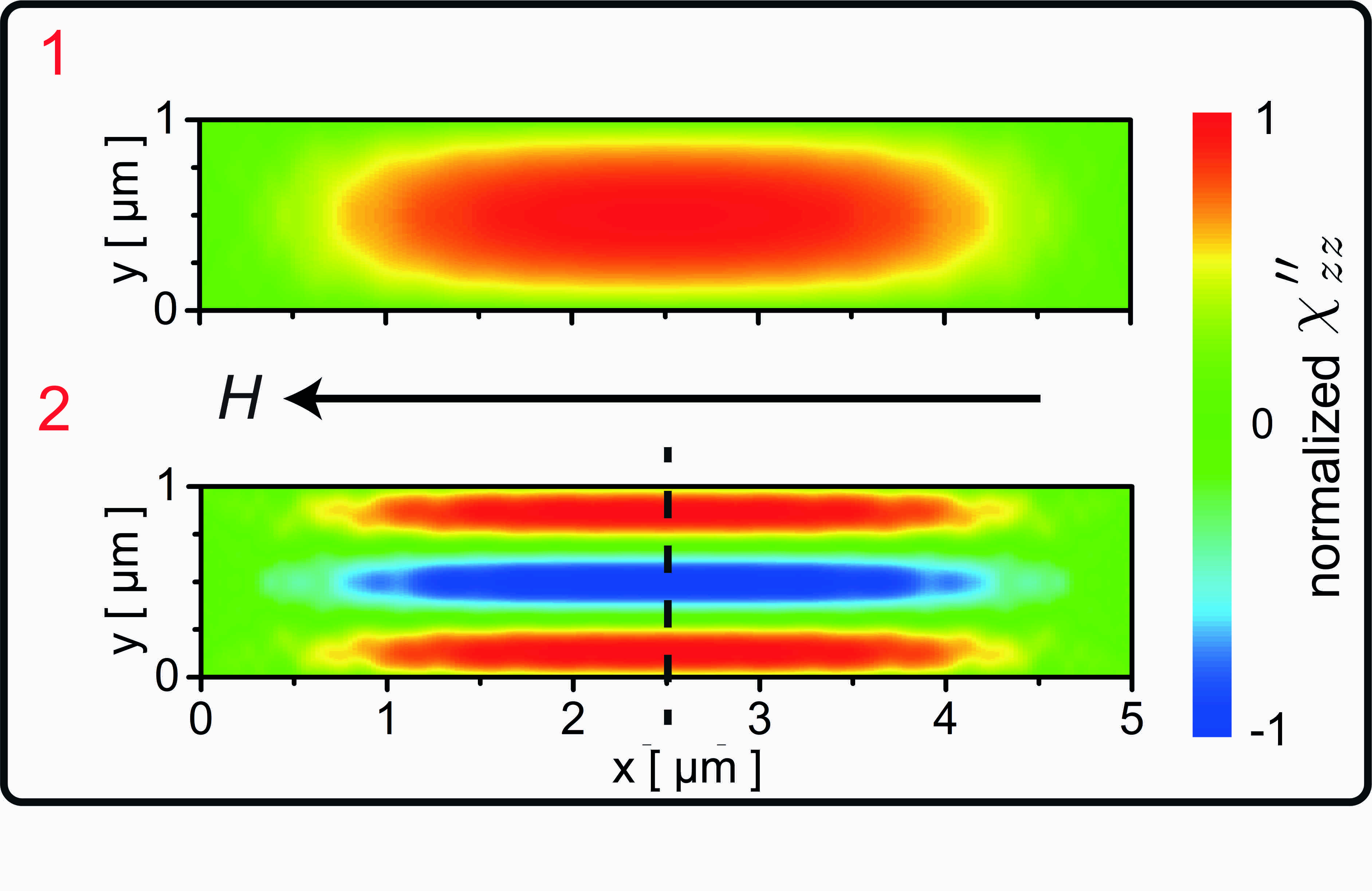}
\caption{Normalized imaginary part of the susceptibility for the magnetic resonances labeled as \textcolor[rgb]{1,0,0}{1} and \textcolor[rgb]{1,0,0}{2} respectively. Resonance \textcolor[rgb]{1,0,0}{1} shows a localized character (quasi uniform mode) with the strongest excitation in the center region of the stripe. Resonance \textcolor[rgb]{1,0,0}{2} exhibits a harmonic dependence across the width of the stripe as expected for a standing spinwave with a wavelength of $734 \nm$. The profile along the dashed line is shown in fig. \ref{fig5}.} \label{fig4}
\end{figure}

In such a confined system multiple magnetic resonances (labeled from \textcolor[rgb]{1,0,0}{1} to \textcolor[rgb]{1,0,0}{4}) with differing positions, linewidths and intensities occur \cite{Hillebrands2002}. The spatial distribution of the normalized imaginary part of the susceptibility for 
the most intense resonance \textcolor[rgb]{1,0,0}{1} is shown color coded in fig. \ref{fig4}. As can be seen this magnetic resonance exhibits the strongest excitation in the center of the stripe and will here be referred to as a localized quasi-uniform mode.
A different mode character can be observed for the less intense magnetic resonance \textcolor[rgb]{1,0,0}{2} in fig. \ref{fig4}. $\chii''_{zz}$ shows a change in sign across the stripe, two nodal lines and a wavelike varying dependence along the dashed line. 

\begin{figure}[htbp]
\includegraphics[width=0.9\columnwidth]{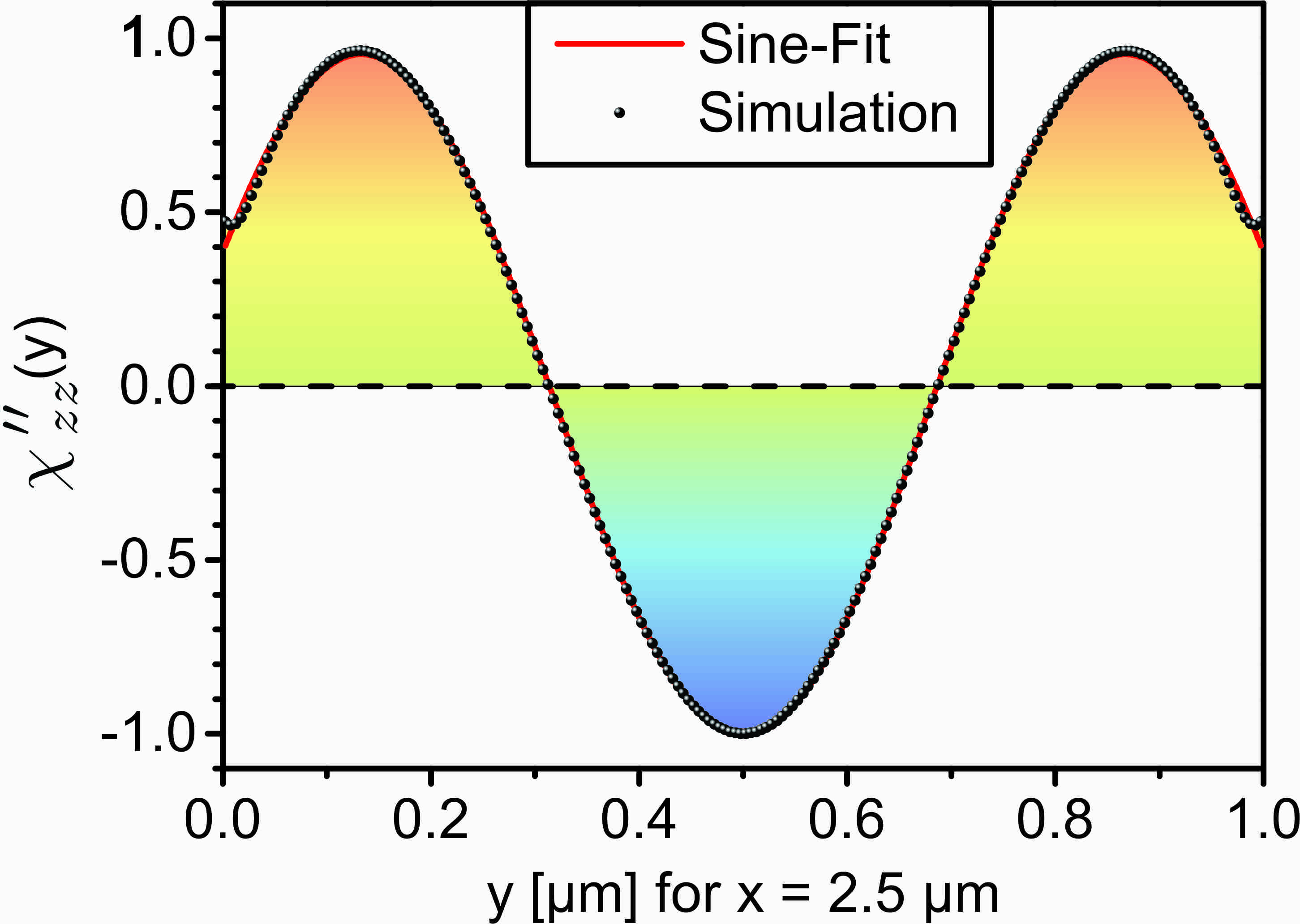}
\caption{Lineprofile of the normalized imaginary part of the susceptibility (black dots) for the magnetic resonances labeled as 2 along the dashed line of fig. \ref{fig4}. The spatial variation is very well approximated by a harmonic dependence of a standing spinwave with a wavelength of $734 \nm$ (red line).} \label{fig5}
\end{figure}

This is very well approximated by a sinusoidal function as shown in fig. \ref{fig5} and resembles the expected modeprofile of a standing spinwave with wavelength $734 \nm$ and dipolar pinning \cite{Guslienko2002} at the edges of the stripe, which deviates from simply closed or open pinning conditions. In a similar consideration the resonances \textcolor[rgb]{1,0,0}{3}, \textcolor[rgb]{1,0,0}{4} can be assigned to higher standing spinwaves across the width of the stripe, where the wavelength decreases for smaller resonance fields.

By such a spatial analysis of the magnetic excitations one can for example explore the dependence of the resonance position, linewidth, mode profile, ellipticity, intensity and pinning conditions on the magnetic parameters as well as on the exact geometry of the magnetic systems. This information can be crucial for planning experiments and lead to a deeper understanding of the multiple resonances in experimentally observed spectra.

\section{Conclusion}\label{Conclusion}

%

In summary, we presented a numerical approach to simulate the FMR of nanosized elements and nanostructured systems using a finite difference method. The simulation utilizes a homogeneous linearly polarized oscillating magnetic field to drive the magnetic system. We would like to point out that this method is very easy to implement and handle. The derivation of simulated FMR spectra, which directly correspond to experimentally detected spectra, is explained in detail. This allows one to compare not only the resonance position, but also the linewidth and intensities of experiment and simulation directly. 

Furthermore the spatial dependence of the magnetic excitations
spin waves and localized resonances can be identified and
visualized. Additionally the time dependent trajectory of the
magnetization as well as the accurate phase relation to the driving field
can be extracted.

\section{Acknowledgments}\label{Acknowledgments}

We acknowledge financial support by the Deutsche Forschungsgemeinschaft (SFB 491) as well as C. Hassel, R. Meckenstock and J.Lindner for fruitful discussions and help in the initial stages of work.

\end{document}